\documentclass{ws-p8-50x6-00}

\def\lapprox{\hbox{\hskip 0.4em\lower 0.7ex\hbox{$\sim$}%
\kern-0.8em\raise 0.45ex\hbox{$<$}\hskip 0.4em}}

\begin{document}

\title{ 
Towards One Tonne Direct WIMP Detectors: Have we got what it takes?
}

\author{R.J. Gaitskell}
\address{Department of Physics and Astronomy, University College
London, Gower St, London WC1E 6BT, UK. \\
E-mail:
rick@gaitskell.com
}

\maketitle

\abstracts{
Experimentally have we got what it takes to pursue the direct observation
of  WIMP interactions down to a sensitivities
of a few events /(100~kg)/year? For a Ge
target  with a low energy threshold ($<20$~keVr) this 
corresponds to a WIMP-nucleon $\sigma \sim 10^{-46} $cm$^{2}$. A number 
of recent 
theoretical papers, making calculations in SUSY-based 
frameworks, show many ($>5$) orders of magnitude spread in 
the possible interaction rates 
for models
consistent with existing Cosmology and Accelerator bounds. 
Some theorists, but certainly not all, are able to generate models, 
that lead to 
interaction rates at the few /kg/day that would be implied by the 
current DAMA annual modulation signal. All theorists 
demonstrate models that generate 
much lower interaction rates. 
This paper takes an unashamed experimentalist`s view of the issues 
that arise when looking 
forward to constructing 1 tonne WIMP detectors.
}

\section{Pursuit of the Grail}

It is very encouraging within the dark matter direct detection field 
to see so many of the technologies proposed over the last decade
coming to fruition. It seems likely that next year will see
a number of different experiments all vying to set the best sensitivity
limits,  and at the same time there will be an overall acceleration in
the rate of progress of the field, as  measured by space carved out in
the log-log exclusion plots (see Fig. \ref{limitplot}). We will
shortly be able to test SUSY models that predict event rates in
the
1/kg/week region.
A common experimental challenge 
at present is how to run a few detector units reliably 
for periods of many months 
in order to collect statistically significant exposures. Of course, some
collaborations are already taking  
long term running for granted. 

It is also worth
taking the time to look at the more distant road ahead, with an 
experimentalist's eye. The challenge for the next generation of
experiments (for ``First Dark'' in 2005-7, say) will be to
 deploy these technologies at the
1 tonne scale.

\begin{figure}[ht]
\begin{center}
\includegraphics [width=0.9\textwidth]{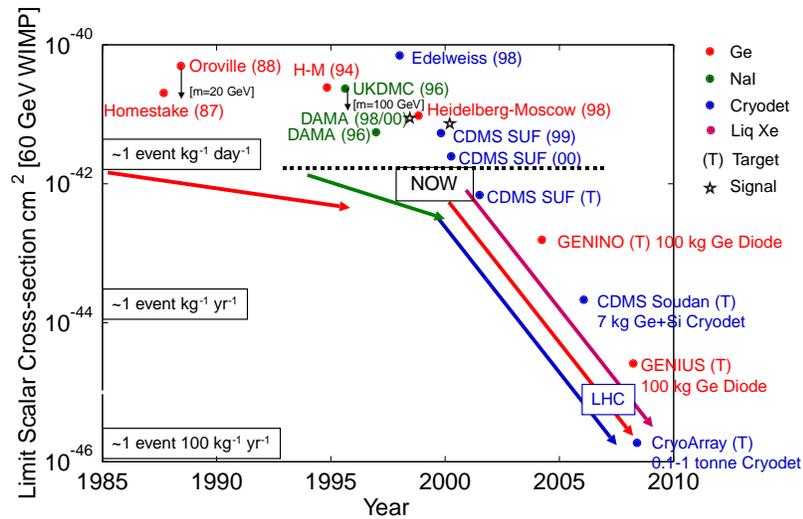}
\end{center}
\caption{
Experiments` 90\% upper CL results for 60 GeV WIMP-nucleon scalar
cross-section versus times of publications. (Not meant to be an
exhaustive selection.) If an experiment performed significantly  better
for a different WIMP mass then this is indicated by the  vertical
arrows. The diagonal arrows are  there to guide
the eye: a gauge of the rate of improvement. The labels in the boxes
give the equivalent event rates in Ge assuming a low recoil threshold
$> 10$~keV. }
\label{limitvstime}
\end{figure}

\subsection{
The Rate of Change of the Rate of Progress of the Field.
}
We should review the past, in order to look at how robust our future 
predictions might be. Figure~\ref{limitvstime} shows the time development of 
the best scalar WIMP-nucleon $\sigma$ limit from the mid 80's to date. 
The first decade 
was dominated by conventional HPGe (and Si) semiconductor detectors. 
The design of these detectors was to some extent `off-the-shelf' and progress was 
achieved, for the most part, by improving the radioactive 
backgrounds around the detectors.
In the mid-90's results from NaI scintillator detectors became 
competitive. They were able to employ pulse shape 
discrimination to make statistical distinctions between 
populations of electron recoil events, and nuclear recoil events. In 
principle, the intrinsic background of the detector and environment 
were no longer the limiting factors, since with sufficient exposure time and 
target mass the limits could be driven down. However, the relatively 
poor quality of the NaI discrimination meant that systematic effects rapidly 
dominate, halting any further improvement with mass$\times$time. The NaI detector 
technology could also be described as `off-the-shelf', however, the
low background and high light yield housing systems were very definitely  
novel. In the case of the DAMA experiment, the deployment of $~100 $kg 
array of NaI also allowed the search for a WIMP annual modulation signal.
At the end of 90's we finally saw results, from new detector technology that had been 
developed specifically for direct detection, take the lead in 
terms of sensitivity.

If we now look forward at some of the predicted goals of a few
experiments over the next decade, it is immediately apparent that the
 forecast rate of progress
appears to be rapidly accelerating. The question is whether this is
simply a `triumph of hope over expectation', or represents a genuinely
improved rate of 
progress that stems from applying detector technologies (2-phase Xe, 
semiconductor and scintillator cryogenic detectors, naked HPGe) that 
were `birthed' with this specific application in mind.

\section{Radioactive Background and Discrimination Goals}

\begin{figure}[!htb]
\begin{center}
\includegraphics [width=.60\textwidth]{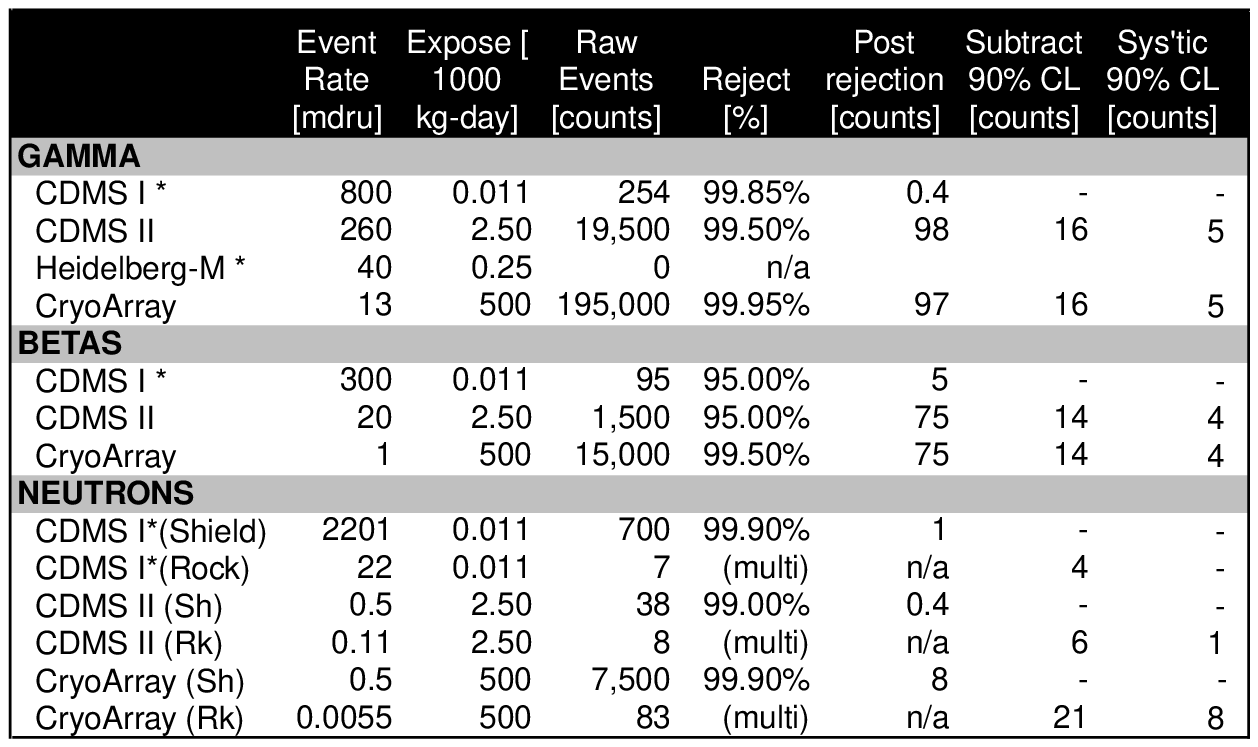}
\includegraphics [width=.353\textwidth]{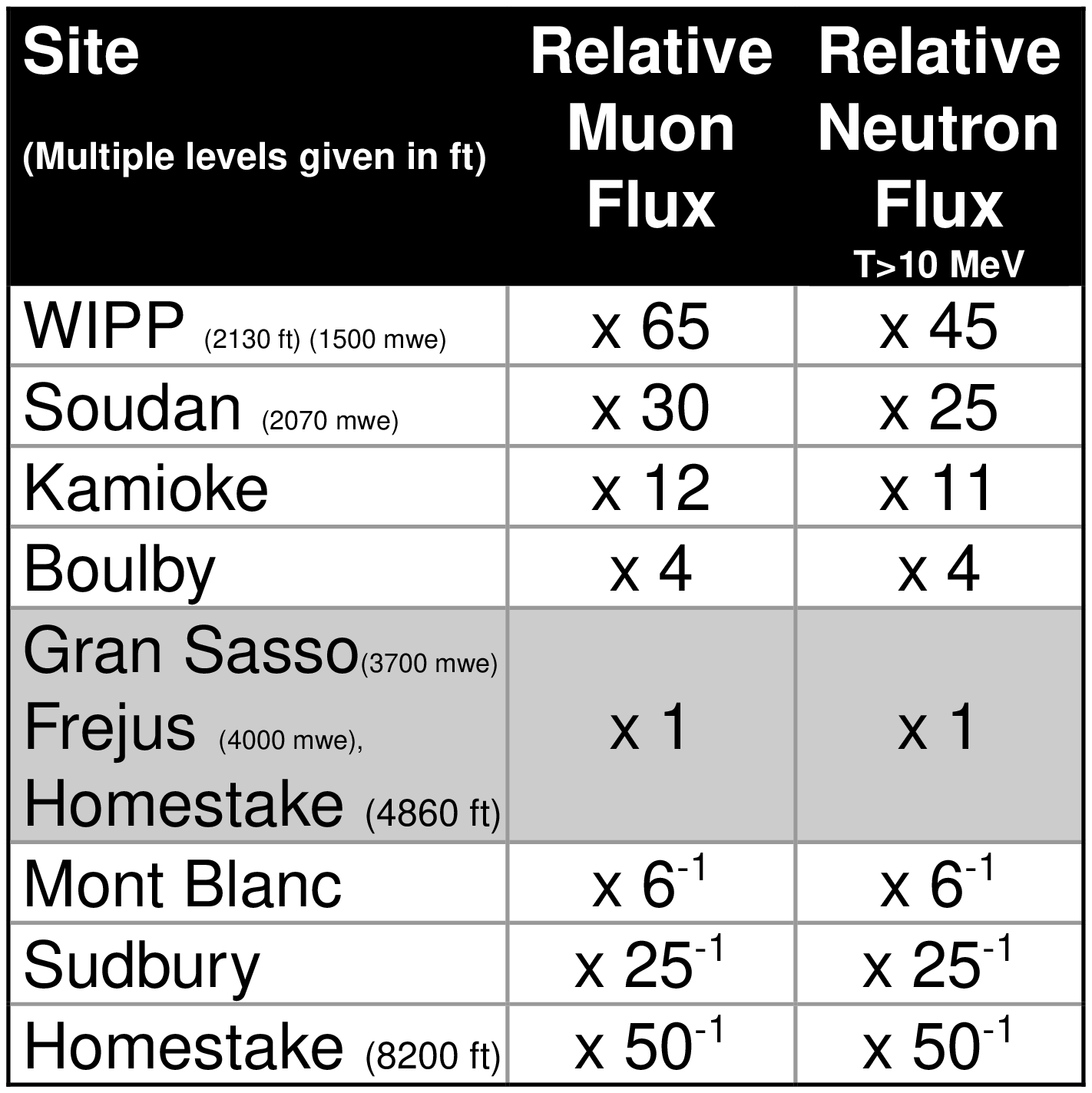}
\end{center}
\caption{
(\textbf{a}) Table summarising the event rates seen, 
or projected, for CDMS I, CDMS II
, Heidelberg-Moscow and CryoArray (see text). This is the
event rate after multiple detector events are vetoed.
The detector material is Ge, and the
energy region is 15-45 keVee for H-M and 15-45 keVr in other 
cases. ``-'' means 
term is negligible, or unimportant for the discussion in this 
paper. For gamma and beta backgrounds the spectra are usually 
flat ($<2\times$) in the energy range considered. For neutrons and 
WIMPs the spectra fall exponentially with slope $\sim 10`s~$keV. In both 
cases we quote the mean value in the range. mdru is $10^{-3}$
/keV/kg/day. The  rows marked with ``*'' are results from actual runs;
the rest are target  performances. 
(\textbf{b}) Table summarising the relative cosmic-ray muon flux, 
and high energy ($T>10$~MeV) neutron flux at a number of deep sites.
Gran Sasso, normalised to 1, is taken at average effective depth of 
3.7~kmwe 
(equiv. to $6.37~10^{-5}$ muons~m$^{-2}$~s$^{-1}$).
} 
\label{deepsites}
\end{figure}

Figure \ref{deepsites}a shows a tabulation of gamma, beta and
neutron background levels and event rates in the energy range 15-45 keV
\footnote{We use 15 keV rather 
than 10 keV to avoid the additional contribution of the 
cosmogenic activation peaks in Ge $\sim$ 10 keV when considering this
simple measure of gamma activity.}
for actual, or projected exposures of the 
CDMS I~\cite{r19prl,gaitskellthisproc}, CDMS II\cite{brinkthisproc},
Heidelberg-Moscow~\cite{hm99} and CryoArray experiments. CryoArray is
the working title for a 1 tonne deployment of semiconductor cryogenic
detectors, of a type similar to that in use in CDMS II. The reason for
breaking up gamma and beta interactions is because the CDMS detectors
have been shown to have different responses to electron-recoils in the
bulk and near the surface.

The first columns of the table show the radioactivity goals and the total
number of events that will be observed during the tabulated exposures for
CDMS I (10.6 kg-days); CDMS II (6.8 kg-years); and CryoArray (1.4
tonne-years). Heidelberg-Moscow is shown as an example of currently
achievable low gamma backgrounds. The later columns show how the
application of detector event by event discrimination can be used first
simply to reject most of the background, but then also to subtract part
of the residual background if the detector response is well enough
understood.~\cite{gaitskelldiscrim} 
The final column shows the systematic error in background
counts associated with a 5\% error in the detector discrimination
calibration. For the case of neutrons the rejection parameter listed
represents the local muon veto efficiency (see below). However, the
 subtraction is performed based on the observed population of
multiple nuclear recoil (NR) events.

The WIMP NR event sensitivity (upper 90\% CL) of CDMS I 
in this limited energy range is $\lapprox 4$ events.
More details of the background levels observed, and the discrimination
performance of the detectors {\it in situ}, can be found
elsewhere~\cite{gaitskellthisproc} and refs. therein.

\begin{figure}
\begin{center}
\includegraphics [width=.27\textwidth]{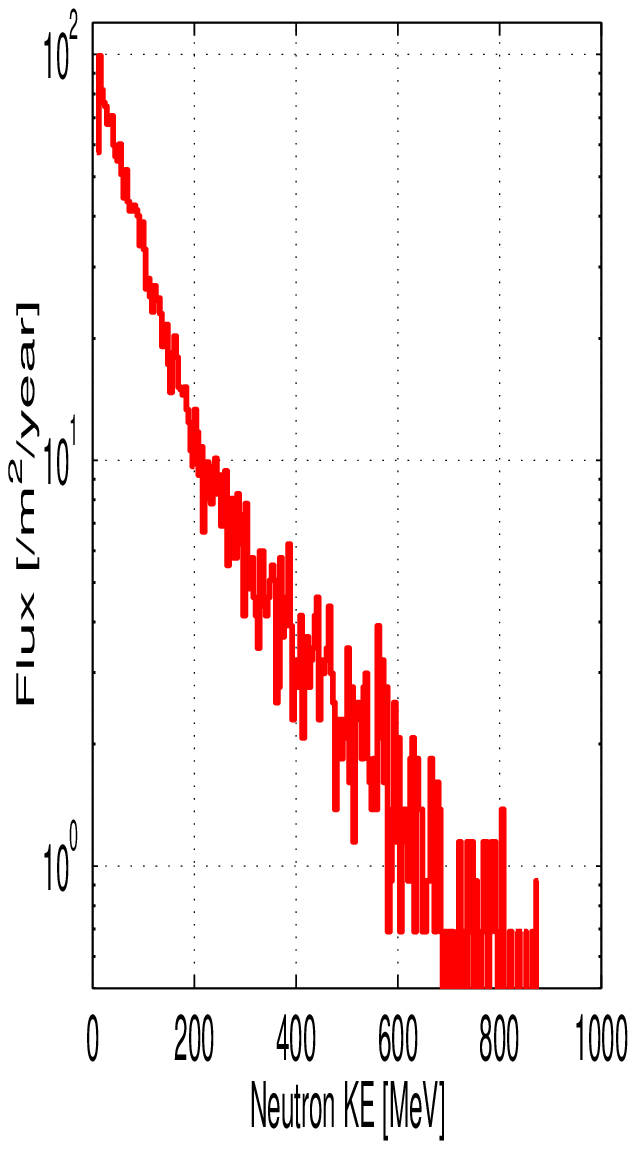}
\includegraphics [width=.26\textwidth]{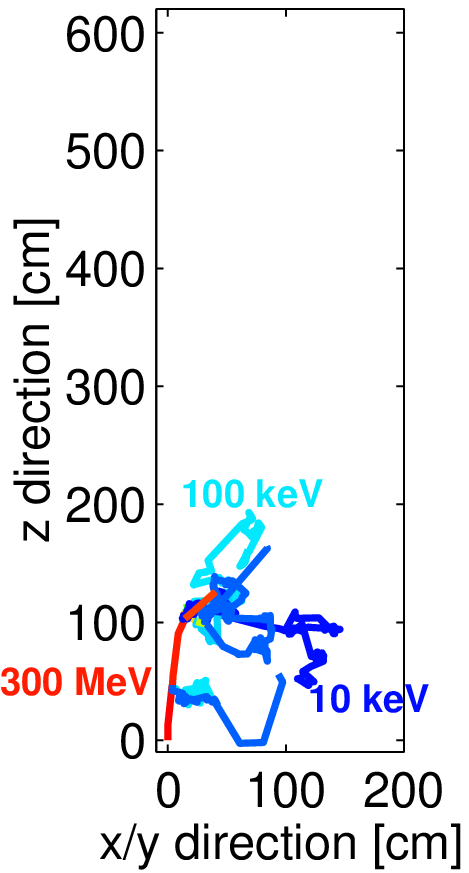}
\includegraphics [width=.26\textwidth]{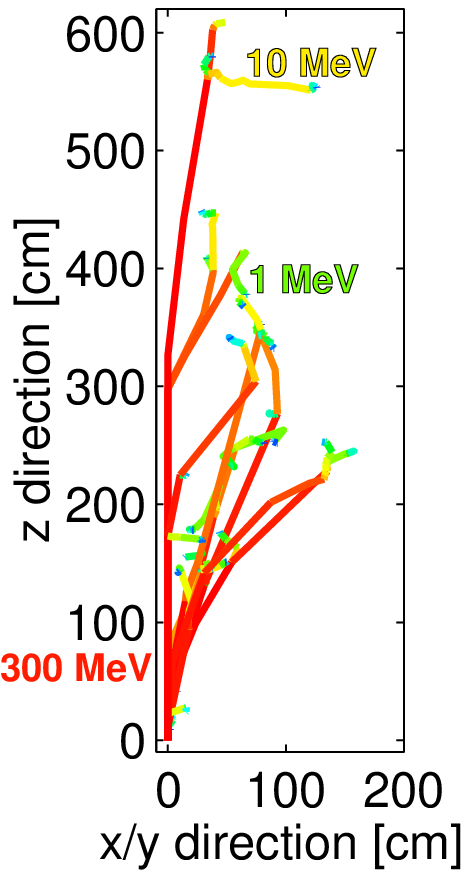}
\end{center}
\caption{
(\textbf{a}) Monte Carlo prediction of high energy neutron flux from the
rock walls arising from muons at a 
depth of 2000 mwe. 
Also shown in plots on the right are tracks from a Monte Carlo simulation
of 300 MeV Neutron(s)  entering shielding material vertically upwards on
the plots. (\textbf{b}) Fe Shield (similar to Pb): only 1 event is shown
for clarity.  (\textbf{c}) Water (similar to polyethylene): 10 separate events
 are shown.}
\label{neutrontracks}
\end{figure}

We will discuss the goals for CDMS II, as an introduction to the CryoArray 1 tonne
goals. 
For CDMS II the WIMP sensitivity goal is around 1 WIMP event/100 kg/day.
The  exposure goal is 2500 kg-days and so $\lapprox 25$ events total
would be  present in the background channels. 
In the original event budget this was to be achieved by lowering the
gamma background by ~3 compared to current CDMS I performance with a 
discrimination goal of 99.5\%. Notably, the gamma discrimination
performance goal for CDMS~II has 
already been exceeded by CDMS~I. In addition, studies are underway to identify the
current dominant sources of gamma activity. 
The beta electron background represents  a greater challenge. Current
levels of observed background are at 300  mdru, whereas the CDMS II goal
is 20 mdru. We believe we have identified the  source of some of the beta
contamination in CDMS~I as due to exposure to a leaking calibration 
source during detector testing. 
Again,  the
discrimination performance of the new ZIP detectors (99.7\%
\cite{brinkthisproc}) has exceeded the
original CDMS II targets,  which will provides some buffer if CDMS~II 
doesn`t
reach the absolute beta background level.
The neutrons causing NR events arise from muons interacting inside the 
Pb/Poly/Cu shielding, and also muons interacting in the cavern rock.
\footnote{The low energy neutrons $T<10$ MeV coming from
radioactivity in the rock are trivially stopped by hydrogenous
shielding, and will not be considered in this analysis}
The former can be tagged  using a local active muon veto
directly around the passive shielding.  80\% veto performance will be adequate
at this  depth to ensure that this type of neutron does not
make a significant contribution.  In fact, a veto efficiency of 99\% is
the stated CDMS II goal, and 99.9\%  has been achieved in the CDMS I
veto, so it is unlikely this neutron source will be a
limitation.~\footnote{At deeper  sites than Soudan 2000~mwe the
requirements for local muon veto performance would be even more  relaxed} 
In addition, a high energy neutron flux is generated in the cavern rock by spallation 
processes of $>$TeV muons, and is proportional to the muon flux. 
For the shallower sites 
\footnote{For the sites $<2.5$ kmwe the 
differential muon flux is 
typically flat up to an energy $250 $GeV$/1 $kmwe, and then falls at
higher energies. 
(WIPP is 1.5~kmwe, Soudan 2.0~kmwe.) For sites deeper than $2.5$
kmwe the muon spectral shape remains constant, only the flux varies.~\cite{pdb00}}
the hardening of the 
muon spectrum with depth leads to an additional factor for the neutron 
flux which can be approximated by 
$~($depth$/2.5 $kmwe$)^{0.75}$. Figure~\ref{deepsites}b records the relative 
muon and neutron fluxes at a number of underground sites, 
and Fig.~\ref{neutrontracks}a 
shows the calculated high energy neutron flux in a cavern at
2000~mwe.~\cite{pererapriv}

The high energy neutrons from the rock are a more 
difficult rejection problem, because of the difficulty in stopping them
(see latter in this section). 
With 
the polyethylene and Pb configuration for CDMS II 
(cylindrical geometry 5 cm plastic scin. veto, 40 cm 
poly, 22 cm Pb, 10 cm poly, then the Cu icebox and detectors)
the residual
signal  due to ``punch-through'' neutrons, after vetoing multiple NR events
using  42 detectors, is expected to 
be$~8$ during the 2500 kg-day exposure, based on extensive Monte
Carlo studies.\cite{pererapriv,yellinpriv} 

In order to reach down a further 2-3 orders of magnitude in WIMP event
rate it  will be necessary to instrument 1 tonne target masses. This
detector  would be sensitive to WIMP-nucleon $\sigma~10^{-46}$cm$^{2}$,
which in  low threshold Ge detectors corresponds to a few events /100
kg/year.  A set of 
possible background contamination and discrimination goals are 
shown in the CryoArray entries of Fig.~\ref{deepsites}a. These are obviously
meant to aid  discussion and are not set in stone. Broadly, these numbers
would reflect a factor 20  improvement in backgrounds, and a factor 10
improvement in discrimination, over the goals for CDMS II. 

In CryoArray the 
gamma background goal would be 13 mdru. This is only
$3\times$ lower than that currently achieved by  Heidelberg in HPGe detectors in the same
energy range. The  discrimination would need to be 99.95\% which again is
only factor 
$3\times$ better than that which is already demonstrated in CDMS I.

The 
situation for beta background is more difficult to make reliable projections for. 
Ideally, by the time the semiconductor cryogenic detectors would be
deployed for the  CryoArray, the performance for gamma and beta
discrimination would be very similar. The best ZIP beta discrimination to date is
$>$99.7\%, and we look forward to further improvements as the signal-to-noise of the
detectors improves further. A beta discrimination of 99.5\% would still
require $\sim$1~mdru which is over two orders of magnitude better than 
current levels of beta contamination. A significant reduction of this
type is a much tougher promise to keep, and  so we would realistically
expect to beat this by some mixture of improved discrimination
(comparable  to that for gammas) and background reduction. 

The neutron situation for CryoArray is interesting. The NR signal for the neutrons from the
shield  would now require veto performance comparable to that
for  CDMS I at 99.9\%, if this experiment was sited at Soudan
(2000~mwe). A  move to Gran Sasso (or equiv at ~4 kmwe) would provide a
neutron  reduction of  ~25x (see Fig \ref{deepsites}b) and the active
muon veto requirement would  then become very modest. The
`punch-through' neutrons at the Soudan site would be more of an
issue. Figure \ref{deepsites}a assumes a factor 20  reduction in the
raw rate of these neutrons for CryoArray in order to meet the target. 
This could be achieve in a number of ways. Firstly, the probability
of  multiple scattering and therefore vetoing neutrons is higher in a
1 tonne  detector vs $\sim10$ kg. This factor needs to be Monte
Carlo'd in detail  since it will depend on the  detailed  composition
of the detector region. Secondly, a move to  a depth of 4 kmwe
would  seem to achieve a suitable reduction. 
Lastly, if the goal was to
perform the experiment at a site at $\lapprox 2000$ mwe 
(e.g. WIPP, or Soudan) then either, a much thicker liquid scintillator
buffer would be required around the detector in order to tag high energy neutrons, or
the cavern rock itself (or an outer heavy shield) would need to be instrumented  with
additional veto detectors in order to catch some part of the  shower
associated with the muon that generated the neutron. Studies in  this
direction are being pursued.~\cite{pererapriv}
 
Figure \ref{neutrontracks}a shows a Monte Carlo simulation of the
high energy neutron flux from the walls of the Soudan cavern at 2000 mwe.
Figures
\ref{neutrontracks}b and \ref{neutrontracks}c show simulations 
of 300 MeV neutron(s) entering
Fe and water, respectively, from the bottom of the plot, in a +ve
$z-$direction. For Fe only 1 event is shown for clarity. The multiplicity
of neutrons generated from the initial scatters is high, typically
$\sim20$, as these processes are predominantly inelastic. The resulting
neutrons have T$\sim 1$MeV and then `diffuse' within the Fe with only
small energy losses per elastic scatter. Based on Monte Carlo
simulations, 100(200)~cm of Fe is required to reduce the flux of neutrons
 by a factor 10, counting those left above 1~MeV(100~keV). For
water 10 separate events
 are shown. In this case the  typical neutron multiplicity 
of the event is lower, being
from 1 to a few. 460~cm of water is required to reduce the flux of
initial neutrons by factor 10 ($T >$ 10 keV). Given the topology
of the interactions discussed above it should
be apparent why a shield containing interleaving Pb and poly would be most
effective at stopping high energy neutrons from ultimately creating lower
energy nuclear recoils (10-100 keV) at the detectors.

According to the Monte Carlo simulations, neutrons in the range $T=50-600
$MeV make equal contributions to the ``punch-through'' neutron population
within the poly/Pb shield of CDMS II. This is because although
the muon-induced neutron flux falls at higher energies, the
effective penetration length of the more energetic neutrons rises almost
exactly to compensate. Below 20 MeV the neutrons are very successfully
moderated by the polyethylene.


\section{
Comparison of Detector Technologies \\and Threshold Behaviour 
}

It is not possible in the space of this paper to comprehensively
 discuss
the technology of the possible detectors for 1
tonne experiments. These proceedings will contain a myriad of technical
details.  However a few observations are appropriate here.

%
\begin{figure}
\begin{center}
\includegraphics [width=.6\textwidth]{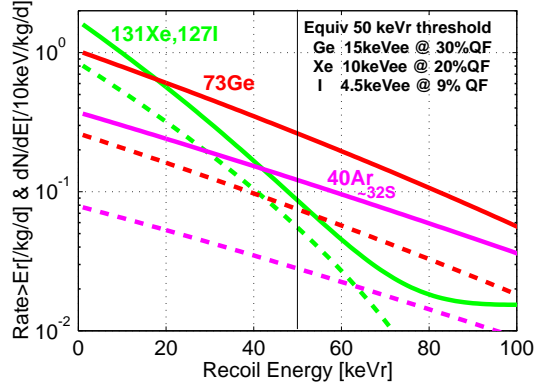}
\end{center}
\caption{Expected rates in both integrated ({\it solid}\,) 
above threshold, and 
differential ({\it dashed}\,) for 300 GeV WIMP with a WIMP-nucleon 
$\sigma=5~10^{-42}~cm^{2}$ The effect of both nuclear coherence 
$\sim A^{2}$
and 
the Form Factor suppression for $q^{2}>0$, as well as kinematics are 
taken into account.
}
\label{formfactor}
\end{figure}

The use of naked HPGe detectors in liquid N$_2$ for the GENIUS/GENINO
experiment is  innovative, and they may be relatively simple to deploy. 
However, if such a system is to be used to probe
$\sigma \sim 10^{-45} $cm$^{2}$ this would require 
$\sim 3~10^{3}$ reduction in the low energy gamma/beta backgrounds of the 
detector assembly compared to the current Heidelberg-Moscow levels, since
the detector has no background discrimination. Low energy (E$<100$ keV)
backgrounds are  very difficult to Monte Carlo reliably since the
observed rate 
will probably be limited by small localized contamination rather than 
distributed levels of U/Th/K. It is also worth noting that the activity
from 
cosmogenically produced $^3$H in Ge, would begin to exceed the GENIUS
target background (0.03 mdru) in the range 0-19~keVee (0-60~keVrecoil)
after only a few hours exposure at sea level. This activity would take
too long to cool, subsequently, when underground.

The three main technologies, in use at kg scales, that exploit nuclear
recoil (NR) discrimination are 2-phase Xe and two types of cryogenic
detectors.~\footnote{
Obviously, NaI scintillators have been very successful to date.
However, the intrinsically weak power of discrimination means that
even based on improvements to 12 p.e./keV, (DAMA currently have 5-7
p.e./keV, UKDM previously at around half this, but this has improved
significantly with new unencapsulated crystals) systematics would still
dominate exposures with sensitivities below $\sigma \sim
10^{-43} $cm$^{2}$.} 

The favoured mode of operation for liquid Xe based detectors involves
measuring both primary scintillation light and drifted ionisation signal
from the interaction. Amplification of the latter signal in the gas phase
means that the main signal-to-noise limitation is in detecting
the light from the primary scintillation. This is likely to determine the
effective energy threshold for the detector. It is too early in the
prototype development of the 2-phase Xe to get an accurate indication of
this threshold value. The importance of the threshold on the
observed rate of interactions is shown in Fig.~\ref{formfactor}.
Clearly the liquid Xe target will scale relatively quickly, since
large liquid noble element calorimeters are already deployed in a
number of particle physics applications, albeit with much higher energy
thresholds.

Cryogenic detectors use the phonon signal combined with either
electron-hole signal in semiconductor crystals, or photon signal in
scintillating crystals.  In contrast to Xe, the signal-to-noise in the
cryogenic detectors is extremely favourable for WIMP detection. In CDMS
the resolution in the ZIP phonon and charge channels is $\lapprox 1~$keV.
This means that high quality event-by-event discrimination occurs above a
low threshold of 10 keVr. However, the challenge in constructing a 1 tonne
array will be to establish the necessary mass production techniques for
$\sim 1000$ detectors. The infrastructure build for CDMS II (42 detectors,
$\sim$10~kg) has lead to the development of cold and warm electronics
systems that already lend themselves well to mass production. Other
cryogenic detector groups are also looking at the same
scaling issues. 

While it is the case that gas TPCs provide elegant methods for
background discrimination, the effective target mass for 1 m$^3$ of gas at
40~torr is 2.3~g$\times$RMM, so $\sim$10,000 m$^3$ of chambers are
required to achieve 1 tonne. The possibility of
increasing the gas operating pressure is being investigated although this
may come at the expense of discrimination quality. 


\begin{figure}
\begin{center}
\includegraphics [width=.72\textwidth]{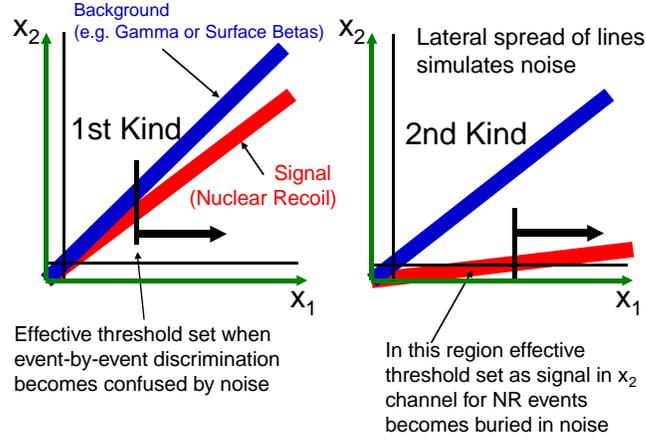}
\end{center}
\caption{
The plots represent background and signal event distributions in
discriminating detectors, which use the ratio of parameters $x_1$ and
$x_2$ to characterise the event. The threshold effect arising
from the relative ratios of the discrimination parameters are
shown in (\textbf{a}) Threshold of First Kind: Signal and background
distributions become confused.
(\textbf{b}) Threshold of Second Kind:
$x_2$ falls into noise for signal events.
}
\label{threshold}
\end{figure}

Figure \ref{threshold}a shows a threshold effect of the ``first kind'',
which arises because the difference in the $x_2/x_1$ ratio for background
and signal is relatively small. A region of confusion occurs for the
event by event discrimination as the  signal-to-noise falls where the
distributions start to overlap. For CDMS $x_1$ is recoil energy and $x_2$
ionisation energy. In Ge~(Si) if $x_2/x_1$ for gamma events is 1, then
$x_2/x_1$ for NR is $\sim1/3~(1/2)$. The confusion threshold in CDMS~I was
below 10 keVr  and  did not
have a significant impact on the sensitivity analyses of the latest
results.     

The CRESST experiment helps illustrate the 
``second kind'' of threshold in Fig. \ref{threshold}b. The intrinsic
separation of lines for electron recoil and signal (NR)
events is significantly greater at 1/7.4, (compared to 1/3 for Ge,
in CDMS). In this case $x_2$ is the amount of
scintillation light from the events, with $x_1$ as the recoil energy
in phonons. This threshold is more subtle, since for smaller
signal-to-noise, the problem is not one of confusion of NR signal with
conventional background events. The small NR events now contain
information in only the $x_1$ (phonon) channel with
$x_2$ (scintillation) lost in noise. This category of $x_1$ only events
could be subject to other forms of fake background such as the
``crack-o-phonics'' that have been seen to emanate from crystal mounts. The
current numbers for the 300~g CaWO$_4$ set up (given that 0.68\% of gamma
event signal is collected as scintillation energy) are such that a 25 keVr
neutron(gamma) would deposit 23~eV(170~eV) in the scintillation light
detector. Such a neutron scintillation signal would likely to be in the
noise.\footnote{
It should be noted that the calibration provided by a
radioactive neutron  source is dominated by O recoils (of the CaWO$_4$
target) for events $E_r > 10 $keV. This is a consequence of the kinematic factor $4 (m_1
m_2)/(m_1+m_2)^2$ in the recoil energy expression, which pushes W recoils to very low energies.
In the case of a WIMP search with CaWO$_4$ the  interaction rate will be dominated by W.
The WIMP mass is better matched to the nuclear mass, and the prospect of
significant $A^2$ enhancement (A=184 for W) seems attractive. However,
for WIMP scattering at a threshold of say, $>32 $keVr, the loss of 
coherence, reflected in the form factor, is such
that the observed rate will be 1/10~th of that for a zero threshold. In
addition, the W (and Ca) quenching factors for NR have not yet been
measured directly, and so it is not known whether they are the same as
that for O nuclei. This is an important parameter to determine in order
to estimate the ultimate threshold and so effective mass of 
the experiment. }
 
\section{
Conclusion
}

\begin{figure}
\begin{center}
\includegraphics [width=.5\textwidth]{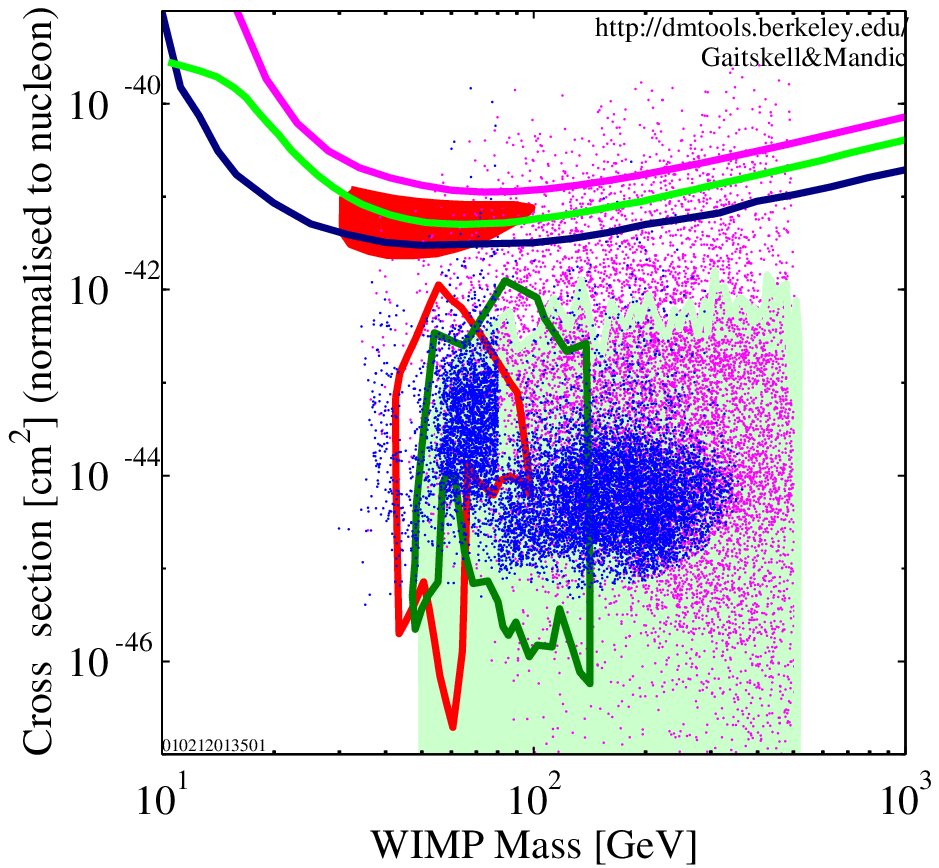}
\includegraphics [width=.44\textwidth]{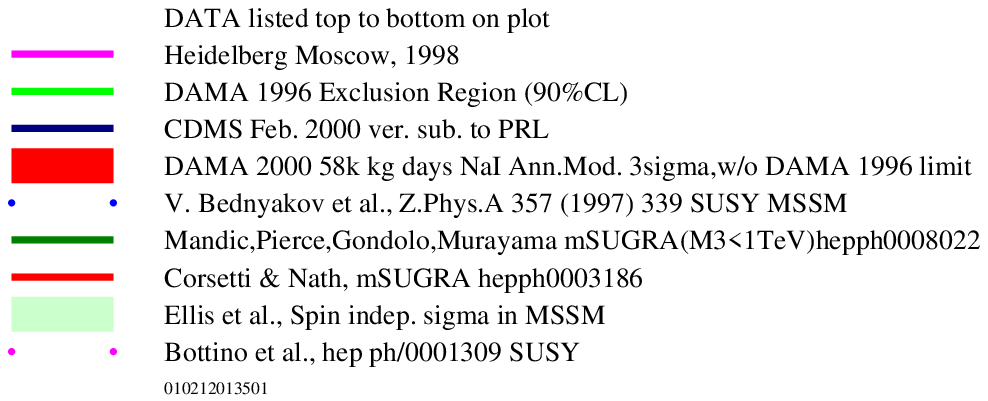}
\end{center}
\caption{Spin-independent WIMP-nucleon cross section vs WIMP mass. The legend on the right
contains labels identifying data sets. Further data can be obtained at 
http://dmtools.berkeley.edu/.
 }
\label{limitplot}
\end{figure}

Part of the recent need to consider larger dark matter 
detectors has arisen because more
sophisticated SUSY based calculations \footnote{The simple crossing
symmetry argument, for evaluating nucleon cross-section, from the
annihilation cross-section, is not satisfactory} are being
performed,~\cite{SUSYall} many of which have driven possible 
 cross-sections down significantly relative to current 
sensitivities (see Fig.~\ref{limitplot}).
 The other
factor being improvements in  current experimental limits, of course!

Experimentalists are often urged to ignore the predictions of
theoreticians, and to hunt for new particles regardless, by the means
that are at hand. While precedent indicates that it may be prudent not to
put too much weight on many theoretical machinations, it is an important
feature of WIMPs that they are motivated by both Particle Physics and
Cosmology. Unfortunately, within SUSY the link between the
WIMP annihilation $\sigma$ and the WIMP-quark $\sigma$ is fairly
loose, and while the former is determined 
within $\sim 10$ by comological bounds ($\Omega_m $) the latter ranges over
many orders of magnitude. If the correlation becomes extremely soft then
one could argue that direct detection is the wrong method to use to look
for Cold Dark Matter (CDM) WIMPs and that a technique more closely
dependent directly on the annihilation
$\sigma$ should be preferred. (However, beware of suppressed branching
ratios if one is detecting a specific decay product.)

If we see SUSY at accelerators within this decade then it is hoped that
enough parameters will be determined to allow calculation of the
LSP properties to see if it can be CDM and, if so, what the
quark interaction rate will be. If we fail to observe SUSY at LHC at the end of this decade it
would seem reasonable to abandon direct detection searches in the absence
of any new compelling framework in which to calculate the WIMP interaction
rate.

In the meantime, it remains a tantalizing possibility 
that one (or more) of
the current experiments may observe an unequivocal
signature for SUSY WIMPs (corresponding to an interaction rate 
that is at the upper end of the theoretically allowed range), 
and thereby
provide a single answer to two of the more entertaining riddles in
Particle Physics and Cosmology.

\end{document}